\documentclass[12pt,a4paper]{article}

\usepackage{amsmath}
\usepackage{amssymb}
\usepackage{subfigure}
\usepackage{epsfig}
\usepackage{t1enc}
\usepackage{cite}

\newcommand{\V}[1]{V_{#1}^{\phantom{\ast}}}
\newcommand{\Vc}[1]{V_{#1}^{\ast}}

\newcommand{\bsmix}{B^0_{s}-\bar B^0_{s}}
\newcommand{\dmix}{D^0-\bar D^0}

\begin{document}
\vspace*{-3cm}
\begin{flushright}
hep-ph/0406151 \\
June 2004
\end{flushright}

\begin{center}
\begin{Large}
{\bf The size of $\boldsymbol{\chi \equiv \arg
(- V_{ts} V_{tb}^* V_{cs}^* V_{cb} )}$ and physics \\[0.2cm]
beyond the Standard Model}
\end{Large}

\vspace{0.5cm}
J. A. Aguilar--Saavedra $^a$, F.J. Botella $^b$, G. C. Branco $^a$
and M. Nebot $^b$ \\[0.2cm] 
{\it $^a$ Departamento de Física and CFTP, \\
  Instituto Superior Técnico, P-1049-001 Lisboa, Portugal} \\
{\it $^b$ Departament de Física Te\`orica and IFIC\\
Universitat de Val\`encia-CSIC, E-46100, Burjassot, Spain}
\end{center}

\begin{abstract}
We analyse the allowed range of values of $\chi$, both in the Standard Model and
in models with New Physics, pointing out that a relatively large value of
$\chi$, e.g. of order $\lambda$, is only possible in models where the unitarity
of the $3 \times 3$ Cabibbo-Kobayashi-Maskawa matrix is violated through the
introduction of extra $Q=2/3$ quarks. We study the interesting case where the
extra quark is an isosinglet, determining the allowed range for $\chi$ and
the effect of a large $\chi$ on various low-energy observables,
such as CP asymmetries in $B$ meson decays. We also discuss the correlated
effects which would be
observable at high energy colliders, like decays $t \to cZ$, modifications of
the cross section and forward-backward asymmetry in $e^+ e^- \to t \bar t$ and
the direct production of a new quark.
\end{abstract}

\section{Introduction}
\label{sec:intro}

The experimental determination of the physical CP-violating phases entering
the quark mixing matrix is of great importance for the study of CP breaking,
providing at the same time stringent tests of the Standard Model (SM).
The Cabibbo-Kobayashi-Maskawa (CKM) matrix \cite{ckm} $V_{3 \times 3}$
describing the
mixing among the known quarks contains nine moduli and
four linearly independent rephasing invariant phases, which can be taken as
\cite{alek,libro}
\begin{align}
\beta = \arg ( - V_{cd} V_{cb}^* V_{td}^* V_{tb}) \,, \quad &
\gamma = \arg ( - V_{ud} V_{ub}^* V_{cd}^* V_{cb}) \,, \nonumber \\ 
\chi = \arg ( - V_{ts} V_{tb}^* V_{cs}^* V_{cb}) \,, \quad &
\chi' = \arg ( - V_{cd} V_{cs}^* V_{ud}^* V_{us}) \,.
\label{phase1}
\end{align}
The phases $\beta$ and $\gamma$ appear in the well-known $(d,b)$
unitarity triangle corresponding to the orthogonality of the first and third
columns of $V_{3 \times 3}$, while $\chi$ and $\chi'$ appear in
other less studied unitarity triangles. The phases $\chi$ and $\chi'$ are
fundamental parameters of $V_{3 \times 3}$ as important as $\gamma$ and 
$\beta$, playing a crucial r\^{o}le in the orthogonality between the $(2,3)$
and $(1,2)$ rows, respectively \cite{silva}.

Within the three-generation SM, the nine moduli and four rephasing-invariant
phases are connected by unitarity, which leads to a series of relations among
these measurable quantities. Such relations provide excellent tests of the SM
\cite{bbnr}, which complement the usual fit
of the unitarity triangle, and have the potential for
discovering New Physics. In the context of the SM, the values of $\chi$ and
$\chi'$ are very constrained and therefore the determination of these phases
provides, by itself, a good test of the SM.

In SM extensions which enlarge the quark sector, the $3 \times 3$ CKM matrix is
a submatrix of a larger matrix $V$. Independently of whether extra
quarks are present or not, one can always choose, without loss of generality, a
phase convention such that \cite{libro}
\begin{equation}
\arg V = \left( 
\begin{array}{cccc}
0 & \chi ^{\prime } & -\gamma & \cdots \\ 
\pi & 0 & 0 & \cdots \\ 
-\beta & \pi +\chi & 0 & \cdots \\ 
\vdots & \vdots & \vdots & \ddots
\end{array}
\right) \,,
\label{phasesCKM}
\end{equation}
which explicitly shows that in the $3 \times 3$ submatrix $V_{3 \times 3}$
only the four phases in Eq.~(\ref{phase1}) are linearly independent.
However, when extra quarks are present the $3 \times 3$ unitarity relations
do not hold, and as a result the range of allowed values for $\chi$ and $\chi'$
may differ from the range implied by the SM. We will show that even in the case
that $3 \times 3$ unitarity does not apply, $\chi'$ is constrained to be rather
small. Therefore, we will concentrate most of our attention on $\chi$,
investigating its expected size within the SM as well as in models with New
Physics.
In Section \ref{sec:2} we use extended unitarity relations to estimate the size
of $\chi$, $\chi'$ within the SM and its extensions, including both the cases
where $3 \times 3$ CKM unitarity is respected and where it is violated.
In Section~\ref{sec:3} a more precise analysis of the range of
variation of $\chi$ in a model with an
extra up singlet is carried out. The effects of a large $\chi$ in some low
energy
observables are examined in Section~\ref{sec:4}, while the effects at high
energy are discussed in Section~\ref{sec:5}. In Section~\ref{sec:6} we draw our
conclusions.

\section{The size of $\boldsymbol{\chi}$ and $\boldsymbol{\chi'}$ in the SM and
its extensions}
\label{sec:2}

It is well known that $\chi ^{\prime }$ has to be very small in the context
of the SM and its extensions which keep the unitarity of the $3 \times 3$ CKM
matrix. This can be seen, for example, using the relation \cite{bbnr}
\begin{equation}
\sin \chi ^{\prime }=\frac{\left| V_{ub}\right| \left| V_{cb}\right| }{%
\left| V_{us}\right| \left| V_{cs}\right| }\sin \gamma \,,
\label{chiprime1}
\end{equation}
which shows that $|\chi'| \lesssim \lambda ^{4}$. Within the SM, the 90\%
confidence level (CL) interval for $\chi'$ is
\begin{equation}
4.95 \times 10^{-4} \leq \chi' \leq 6.99 \times 10^{-4} \quad \mathrm{(SM)} \,.
\end{equation}
This range is obtained with a fit to the measured CKM matrix
elements in Table~\ref{tab:1}, together with $\varepsilon$, the $B^0$ mass
difference and the time-dependent CP asymmetry in $B_d^0 \to \psi \, K_S$,
$S_{\psi K_S}$, all collected in Table~\ref{tab:2} (see Refs.~\cite{pdb,hfag}).

\begin{table}[htb]
\begin{center}
\begin{tabular}{cc}
Element & Exp. value \\
\hline
$|V_{ud}|$ & $0.9734 \pm 0.0008$ \\
$|V_{us}|$ & $0.2196 \pm 0.0026$ \\
$|V_{ub}|$ & $0.0036 \pm 0.0010$ \\
$|V_{cd}|$ & $0.224 \pm 0.016$ \\
$|V_{cs}|$ & $0.989 \pm 0.014$ \\
$|V_{cb}|$ & $0.0402 \pm 0.0019$
\end{tabular}
\caption{Experimental values of CKM matrix elements.}
\label{tab:1}
\end{center}
\end{table}

\begin{table}[htb]
\begin{center}
\begin{tabular}{cc}
 & Exp. value \\
\hline
$\varepsilon$ & $(2.282 \pm 0.017) \times 10^{-3}$ \\
$\delta m_{B_d}$ & $0.489 \pm 0.008 ~\mathrm{ps}^{-1}$ \\
$S_{\psi K_S}$ & $0.734 \pm 0.054$
\end{tabular}
\caption{Additional observables required for the fit of the CKM matrix.}
\label{tab:2}
\end{center}
\end{table}

Even in models where $V_{3 \times 3}$ is not unitary, but part of a larger
unitary matrix $V$,
$\chi ^{\prime }$ is constrained to be rather small \cite{libro}.
From orthogonality of the
first two columns of $V$, one readily obtains
\begin{equation}
\cos \chi ^{\prime } \geq \frac{ |V_{ud}|^2 + |V_{cs}|^2 + |V_{us}|^2
+ |V_{cd}|^2 - |V_{ud}|^2 |V_{cs}|^2 - |V_{us}|^2 |V_{cd}|^2 - 1}{2 |V_{ud}|
|V_{us}| |V_{cd}| |V_{cs}|} \,,
\label{chiprime2}
\end{equation}
implying $\cos \chi' \geq 0.9983$ and 
\begin{equation}
|\chi'| \leq 0.0579
\label{chiprime3}
\end{equation}
at 90\% CL. This limit is robust in the presence of New Physics, since
the moduli involved are obtained from experiment
through tree-level decays, where the SM is expected to give the dominant
contribution. From the strict bound of Eq.~(\ref{chiprime3}) it is clear that it
will be very difficult to obtain a direct measurement of $\chi'$. Therefore, in
the remaining of this work we will focus our attention on $\chi$.

Within the SM and any extension where $V_{3 \times 3}$ is unitary, like
supersymmetric or multi Higgs doublet models, we have the relation
\begin{equation}
\sin \chi = \frac{|V_{ub}| |V_{us}|}{|V_{cb}| |V_{cs}|} \sin (\gamma + \chi'
-\chi ) \,,
\label{Chi3}
\end{equation}
which shows that $|\chi| \lesssim \lambda^2$ in any model where $3 \times 3$ CKM
unitarity holds. In particular, within the SM one obtains at 90\% CL
\begin{equation}
0.015 \leq \chi \leq 0.022 \quad \mathrm{(SM)} \,.
\label{Chi2}
\end{equation}
The only models in which $\chi$ can be significantly larger than $\lambda^2$
are those in which $V_{3 \times 3}$ is not unitary, what can only be achieved
by enlarging the quark sector. The most simple way of doing this is with the
introduction of new quark singlets \cite{paconpb,sin}.\footnote{The addition of
a sequential fourth generation is another possibility, but it is disfavoured by
two facts: ({\em i\/}) the experimental value of the oblique correction
parameters only leave a small range for the masses of the new quarks; 
({\em ii\/}) anomaly cancellation requires the introduction of a new lepton
doublet, in which the new neutrino should be very heavy, in contrast with the
small masses of the presently known neutrinos.}
Quark singlets often arise in grand unified theories \cite{unif1,unif2}
and models with extra dimensions at the electroweak scale \cite{paco}. They
have both their left- and right-handed
components transforming as singlets under $\mathrm{SU}(2)_L$, thus their
addition to the SM particle content does not spoil the cancellation of triangle
anomalies.
In these models, the charged and neutral current terms of the Lagrangian in the
mass eigenstate basis are
\begin{eqnarray}
{\mathcal L}_W & = & - \frac{g}{\sqrt 2}
\,\bar u_L \gamma^\mu V d_L \,W_\mu^+ +\mathrm{h.c.} \,, \nonumber \\
{\mathcal L}_Z & = & - \frac{g}{2 c_W} \left(
\bar u_L \gamma^\mu X u_L - \bar d_L \gamma^\mu U d_L
  - 2 s_W^2 J_\mathrm{EM}^\mu \right) Z_\mu \,,
\label{ec:2}
\end{eqnarray}
where $u=(u,c,t,T,\dots)$ and $d=(d,s,b,B,\dots)$, $V$ denotes the extended CKM
matrix and $X = V V^\dagger$,
$U = V^\dagger V$ are hermitian matrices. $X$ and $U$ are not
necessarily diagonal and thus flavour-changing neutral (FCN) couplings exist at
the tree level, although they are naturally suppressed by the ratio of the
standard
quark over the heavy singlet masses \cite{paconpb}. Moreover, the diagonal $Zqq$
couplings, which are given by the
diagonal entries of $X$ and $U$ plus a charge-dependent term, are also modified.
Within the SM $X_{uu} = X_{cc} = X_{tt} = 1$, $X_{qq'} = 0$ for $q \neq q'$,
$U_{dd} = U_{ss} = U_{bb} = 1$ and $U_{qq'} = 0$ for $q \neq q'$.
The addition of up-type $Q=2/3$ singlets modifies the first two of these
equalities, while the addition of down-type $Q=-1/3$ ones modifies the last two.
For our purposes, it is sufficient to consider that either up- or down-type
singlets are added to the SM particle content. We analyse in turn these two
possibilities.

\subsection{Models with down-type singlets}

In this case, and assuming that there are $n_d$ extra down singlets,
the CKM matrix $V$ is a $3 \times (3+n_d)$ matrix 
 consisting of the first three rows of a
$(3 + n_d) \times (3 + n_d)$ unitary matrix, and $X = 1_{3 \times
3}$. From orthogonality of the second
and third columns of $V$, one obtains the generalisation of Eq.~(\ref{Chi3}),
\begin{equation}
\sin \chi =\frac{|V_{ub}| |V_{us}|}{|V_{cb}| |V_{cs}| } \sin ( \gamma +
\chi' - \chi) - \frac{\text{Im} \left(U_{bs} e^{-i \chi} \right)}{|V_{cb}|
|V_{cs}|} \,.
\label{Chi5}
\end{equation}
From the present bound on $b \rightarrow s \ell^+ \ell^-$, one obtains that 
at most $|U_{bs}| \simeq 10^{-3} \sim \lambda^4$ \cite{fits,jaas},
thus implying that in this
class of models $\chi$ cannot be significantly larger than $\lambda^2$.

\subsection{Models with up-type singlets}

In these models the quark mixing matrix is a $(3+n_u) \times 3$ matrix, with
$n_u$ the number of extra singlets, and $U=1_{3 \times 3}$. Almost all the
effects discussed in this paper can
be already obtained in the minimal extension with $n_u=1$, in which case
the quark mixing matrix has dimension $4 \times 3$. From orthogonality of the
second and third columns one obtains the generalisation of Eq.~(\ref{Chi3}) for
this model,
\begin{equation}
\sin \chi = \frac{|V_{ub}| |V_{us}|}{|V_{cb}| |V_{cs}|} \sin ( \gamma + \chi'
- \chi ) + \frac{|V_{Tb}| |V_{Ts}|}{|V_{cb}| |V_{cs}|} \sin (\sigma - \chi) \,,
\label{Chi6}
\end{equation}
where $\sigma \equiv \arg ( \V{Ts}\Vc{Tb}\Vc{cs}\V{cb} )$.
$\chi$ may be of order $\lambda$ if $V_{Ts} \sim \lambda^2$ and 
$V_{Tb} \sim \lambda$, but the possible constraints from FCN currents in the up
sector must also be kept in mind. From orthogonality of the second and third
rows of $V$, one gets
\begin{equation}
\sin \chi = \frac{\mathrm{Im} \; X_{ct}}{|V_{cs}| |V_{ts}|} + O(\lambda^2) \,.
\label{ec:chi-x}
\end{equation}
In contrast with models containing down-type singlets, where the size of all
FCN couplings is very restricted by experiment, present limits on $X_{ct}$
are rather
weak. The most stringent one, $|X_{ct}| \leq 0.41$ with a 95\% CL, is derived
from the non-observation of single top production at LEP, in the process
$e^+ e^- \to t \bar c$ and its charge conjugate \cite{opal}. This bound does
not presently
provide an additional restriction on the size of $\chi$.
In models with extra up singlets $|X_{ct}|$ can be of order $\lambda^3$
\cite{jaas}, yielding $\chi \sim \lambda$. 

From Eq.~(\ref{ec:chi-x}) one derives some important
phenomenological consequences. First, we observe that a sizeable $\chi$ is
associated to
a FCN coupling $X_{ct} \sim 10^{-2}$, which leads to FCN decays $t \to cZ$ at
rates observable at LHC. In addition, the modulus of $X_{ct}$ obeys the
inequality \cite{prl}
\begin{equation}
|X_{ct}|^2 \leq (1-X_{cc}) (1-X_{tt}) \,,
\label{ec:desig}
\end{equation}
which is verified in any SM extension with any number of up- and/or down-type
quark singlets (in
particular, with only one $Q=2/3$ singlet the equality holds). We note that
within the SM, $X_{cc} = X_{tt} = 1$ and hence $X_{ct} = 0$. This relation
shows that necessary conditions (and
sufficient for the case of only one singlet)
for achieving $X_{ct} \sim 10^{-2}$ 
are to have a small deviation $O(\lambda^4)$ of $X_{cc}$ from unity (which is
allowed by the
measurement of $R_c$ and $A_\mathrm{FB}^{0,c}$) and a deviation of
$X_{tt}$ from unity of order $\lambda^2$. The latter could be measured in
$t \bar t$ production
at a future $e^+ e^-$ linear collider like TESLA. There is also a
decrease of $|V_{tb}|$ from its SM value $|V_{tb}| \simeq 0.999$, which is
however harder to detect experimentally, because the expected precision in the
measurement of this quantity at LHC is around $\pm 0.05$ \cite{single}.
Last, but not least, 
this deviation of $X_{tt}$ from unity is only possible if the new quark has a
mass below 1 TeV, in which case it would be directly produced and
observed at LHC.

\section{Detailed analysis of the range of $\boldsymbol{\chi}$ with an extra up
singlet}
\label{sec:3}

The analysis of the previous section has shown that $\chi$ can in principle be
of order $\lambda$ in models with up quark singlets. In order to determine
its precise range of variation, it is mandatory to perform an analysis including
constraints from a variety of processes for which the predictions
are affected by the inclusion of an extra up quark. We summarise here the most
relevant ones.
\begin{enumerate}
\item The presence of the new quark and the deviation of $|V_{tb}|$ and 
$X_{tt} \simeq |V_{tb}|^2$ from the SM predictions yield new
contributions to the oblique parameters $S$, $T$ and $U$. The most
important one corresponds to the $T$ parameter, approximately
\begin{equation}
\Delta T = \frac{N_c}{16 \pi s_W^2 c_W^2} (1-X_{tt}) \left[ -18.4 + 7.8 \log y_T
\right] \,,
\label{ec:deltaT}
\end{equation}
with $N_c = 3$ the number of colours and $y_T = \left( \bar m_T /M_Z \right)^2$.
The present experimental measurement $\Delta T = -0.02 \pm 0.13$ sets stronger
limits on $V_{tb}$ and $X_{tt}$ than the $S$, $U$ parameters or the
forward-backward asymmetry $A_\mathrm{FB}^{(0,b)}$.

\item The deviation of $X_{cc}$ from unity modifies the $Zcc$
couplings and thus the prediction for $R_c$ and the forward-backward
asymmetry $A_\mathrm{FB}^{(0,c)}$. The precise measurement of these quantities
sets a stringent constraint on $X_{cc}$, with a direct influence on $\chi$, as
shown by Eqs.~(\ref{ec:chi-x}), (\ref{ec:desig}).

\item The FCN coupling $X_{uc}$ mediates a tree-level contribution 
to $D^0- \bar D^0$ mixing, which is kept within experimental limits for $X_{uc} 
\lesssim 5 \times 10^{-4}$.

\item The new quark $T$ gives additional loop contributions to
$K$ and $B$ oscillations and rare decays $K^+ \to \pi^+ \nu \bar \nu$,
$K_L \to \mu^+ \mu^-$, $b \to s \gamma$ and $b \to s l^+ l^-$. The new terms
are similar to the top ones, but proportional to some combination of the CKM
matrix elements $V_{Td}$,
$V_{Ts}$, $V_{Tb}$ and with the corresponding Inami-Lim functions evaluated at
$x_T = (\bar m_T/M_Z)^2$. For the unrealistic case $x_T \simeq x_t$ the
Inami-Lim functions for the $t,T$ quarks take similar values, and the sum of
both terms may be very similar to the top SM contribution. Therefore,
in this situation the constraints on $V_{Td}$, $V_{Ts}$, $V_{Tb}$ are rather
loose. However, for $m_T \gtrsim 300$ GeV these observables provide important
constraints on $V_{Td}$ and $V_{Ts}$, forcing also $V_{td}$ and $V_{ts}$ to lie
in their SM range.
\end{enumerate}
These and other less important constraints like $\varepsilon'/\varepsilon$
have been taken into account in our analysis \cite{jaas}. It is important to
note that the most recent bound on the CP asymmetry in $b
\to s \gamma$ \cite{babar} is still not relevant. Using an appropriate
generalisation of the formulas in Refs.~\cite{porod} for the present case, we
always find $|A_\mathrm{CP}^{b \to s \gamma}| \lesssim 0.02$, to be compared
with the experimental 90\% CL interval $-0.06 \leq
A_\mathrm{CP}^{b \to s \gamma} \leq 0.11$.

\begin{figure}[htb]
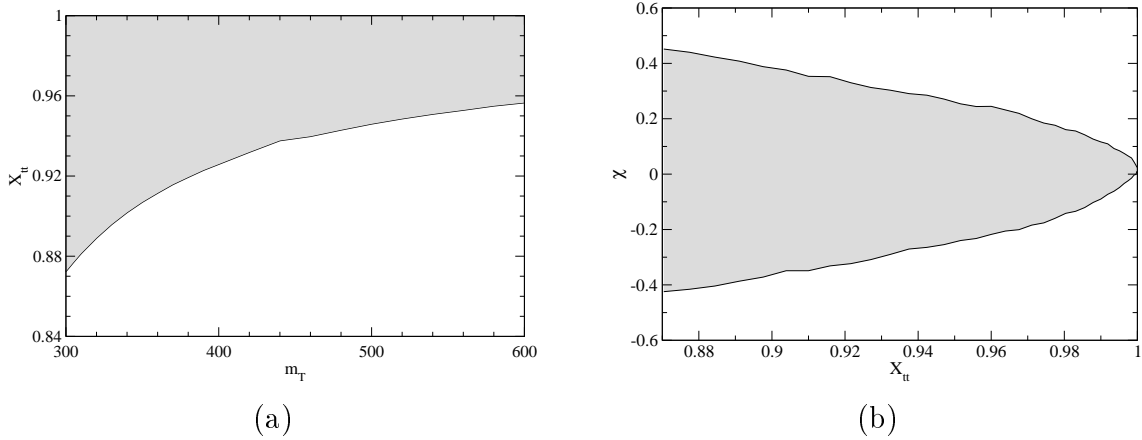

\begin{center}
\begin{tabular}{ccc}
\epsfig{file=Figs/Xtt.eps,width=7.05cm,clip=} & ~ &
\epsfig{file=Figs/X-Xtt.eps,width=7.05cm,clip=} \\
(a) & & (b)
\end{tabular}
\caption{(a) Allowed interval of $X_{tt}$ (shaded area) as a function of the
mass of the new quark (adapted from Ref. \cite{jaas}).
(b) Allowed interval of $\chi$ (shaded area) as a function of $X_{tt}$.}
\label{fig:Xtt-X}
\end{center}
\end{figure}

We will conservatively assume that the mass of the new quark $T$ is of 300 GeV
or larger. Present Tevatron Run II measurements seem to exclude the existence of
a new quark with a mass around 200 GeV and decaying to $Wb$ \cite{tmass}.
However, we will
briefly comment on the situation if the new quark is lighter than 300 GeV.
We remark that
the allowed range of $\chi$ only depends on the mass of the new quark
through the $m_T$ dependence of $X_{tt}$. The possible values of $X_{tt}$ are
constrained mainly by the $T$ parameter, and are shown
in Fig.~\ref{fig:Xtt-X}a as a function of $m_T$. For a
fixed $X_{tt}$, the interval in which $\chi$ can vary turns out to be
independent of $m_T$. The allowed range of $\chi$ as a function of $X_{tt}$ is
plotted in Fig.~\ref{fig:Xtt-X}b. We observe that, as anticipated in the
previous section,
a deviation of $X_{tt}$ from unity is necessary in order to have $\chi$ large.
For $X_{tt} = 1$ the range of $\chi$ reduces to the SM interval (see
Fig.~\ref{fig:Xtt-X}b).

We present two examples of matrices $V$ for $m_T = 300$ GeV which give large
$|\chi|$ with positive and negative sign, respectively. We have not chosen
examples which maximise $|\chi|$ but have instead selected two matrices
which yield theoretical predictions for presently known observables in very
good agreement with experiment,
while showing significant departures in $\chi$ from the SM
expectation. We write the full $4 \times 4$ unitary matrices, although only the 
$4 \times 3$ submatrices enter the charged current interactions. We
choose the phase parameterisation in Eq.~(\ref{phasesCKM}), in which
the values of the four phases in Eq.~(\ref{phase1}) are easy to read directly
from the matrices. The first example is
\begin{align}
\left| V_{300}^{(+)} \right| & = \left( 
\begin{array}{cccc}
0.9748 & 0.2229 & 0.0038 & 0.0097 \\ 
0.2230 & 0.9733 & 0.0406 & 0.0362 \\ 
0.0072 & 0.0355 & 0.9422 & 0.3332 \\ 
0.0009 & 0.0419 & 0.3327 & 0.9421
\end{array}
\right) \,, \nonumber \\
\arg V_{300}^{(+)} & = \left( 
\begin{array}{cccc}
0 & 6.92 \times 10^{-4} & -0.8222 & -0.1046 \\ 
\pi & 0 & 0 & 0 \\ 
-0.4099 & \pi + 0.3513 & 0 & 1.940 \\ 
0 & 2.346 & 0.1001 & -1.106
\end{array}
\right) \,.
\label{ec:matr1}
\end{align}
This matrix has $\beta = 23.5^\circ$, $\gamma = 47.1^\circ$ in the $(d,b)$
unitarity triangle. While $\beta$ is close to the SM prediction, $\chi = 0.35$
presents a large deviation from the SM
value. For this matrix $S_{\psi K_S} = 0.70$, with $\epsilon$, $\delta m_B$,
$\mathrm{Br}(b \to s \gamma)$, $\mathrm{Br}(b \to s l^+ l^-)$ and the rest of
observables considered in Ref.~\cite{jaas} also in good agreement with
experiment. The second example is
\begin{align}
\left| V_{300}^{(-)} \right| & = \left( 
\begin{array}{cccc}
0.9748 & 0.2229 & 0.0038 & 0.0090 \\ 
0.2230 & 0.9733 & 0.0419 & 0.0347 \\ 
0.0077 & 0.0406 & 0.9571 & 0.2865 \\ 
0.0024 & 0.0366 & 0.2864 & 0.9574
\end{array}
\right) \,, \nonumber \\
\arg V_{300}^{(-)} & = \left( 
\begin{array}{cccc}
0 & 5.17 \times 10^{-4} & -1.020 & 0.0700 \\ 
\pi & 0 & 0 & 0 \\ 
-0.3608 & \pi-0.2382 & 0 & -1.576 \\ 
0 & -1.026 & 0.8784 & 2.449
\end{array}
\right) \,.
\label{ec:matr2}
\end{align}
For this matrix $\beta = 20.7^\circ$, $\gamma = 58.4^\circ$ in the $(d,b)$
unitarity triangle and $\chi = -0.24$, in clear contrast with the SM
prediction. We find that
$S_{\psi K_S} = 0.74$, with the other observables agreeing with experimental
data. In both examples we observe that $X_{ct} = -V_{c4} V_{t4}^*$ has a
large imaginary part (in this phase convention), as required for a large
$\chi$ according to Eq.~(\ref{ec:chi-x}). The values obtained for
$\chi$ are of the same order as the estimates given in the previous section.
We stress that $\chi$ can be of order $\lambda$ while keeping $S_{\psi K_S}$
close to its experimental value. Hence, a future improvement of this
measurement (e.g. a reduction of the statistical error by a factor of two)
has little effect on our results.

\section{Low energy observables sensitive to $\boldsymbol{\chi}$}
\label{sec:4}

The decay $B_d^0 \rightarrow \phi K_{S}$ is an interesting example
in which CP-violating effects sensitive to $\chi$ may be found, with the
advantage that $B_d^0$ mesons can be produced at present $B$ factories. The
time-dependent CP asymmetry is given by: 
\begin{equation}
S_{\phi K_S} = \frac{2 \, \text{Im} \, \lambda _{\phi K_{S}}}{1 +
|\lambda_{\phi K_S}|^2} \,,
\end{equation}
where
\begin{equation}
\lambda_{\phi K_{S}} = \left( \frac{q}{p} \right)_{B_d^0} \;
\frac{A(  \bar B_d^0 \rightarrow \phi \, \bar K^0)}{A(
B_d^0 \rightarrow \phi \, K^0)} \;
\left( \frac{q}{p} \right)_{K^0} \,.
\label{ec:18}
\end{equation}
The $q/p$ factors come from $B_{d}^{0}$ and $K^{0}$ mixing. The SM decay
amplitudes are, to a very good approximation, 
\begin{eqnarray}
A( \bar B_d^0 \rightarrow \phi \, \bar K^0) & = &
a(x_t) V_{tb} V_{ts}^* \,,
\nonumber \\
A(B_d^0 \rightarrow \phi \, K^0) & = & a(x_t) V_{tb}^* V_{ts} \,,
\label{ec:19}
\end{eqnarray}
with $a(x_t)$ a function of $x_t = (\overline{m_t}/M_W)^2$, to be specified
later. In the SM, or in any model without New Physics in the
decay amplitudes, $\lambda_{\phi K_S}$ can be
related to
its analogous in the $\psi K_S$ decay channel,
\begin{equation}
\lambda_{\psi K_{S}} = \left( \frac{q}{p} \right)_{B_d^0} \;
\frac{V_{cb} V_{cs}^*}{V_{cb}^* V_{cs}} \;
\left( \frac{q}{p} \right)_{K^0} \,.
\end{equation}
Bearing in mind the definition of $\chi$ we can write
\begin{equation}
\lambda_{\phi K_{S}} = \lambda_{\psi K_{S}} e^{-2 i \chi} \,,
\label{ec:21}
\end{equation}
so that defining $\bar \beta$ by $\lambda_{\psi K_S} = -e^{- 2i \bar \beta}$
($\bar \beta = \beta$ in the SM, but these two angles may differ if there are
new contributions to the mixing) we have
\begin{equation}
S_{\phi K_S} = \sin (2 \bar \beta + 2 \chi) \,.
\end{equation}
Therefore, if a substantial
departure from the approximate SM prediction $S_{\phi K_S} \simeq S_{\psi K_S}$
is
confirmed, it cannot be explained in models with $3 \times 3$ CKM unitarity and
without new contributions to the decay amplitudes.

The best place to measure $\chi$ is in CP asymmetres in $\bsmix$ oscillations
and decay. In the SM the $B_s^0$ mixing factor is
\begin{equation}
\left( \frac{q}{p} \right)_{B_s^0} = \frac{M_{12}^{B_s}}{|M_{12}^{B_s}|} =
\frac{(V_{ts} V_{tb}^*)^2}{|V_{ts} V_{tb}^*|^2} = e^{2 i \chi} \,.
\end{equation}
In any channel without a weak phase in the decay amplitude,
for example in the $D_{s}^{+}D_{s}^{-}$ and $\psi \, \phi$ channels,
the time dependent CP asymmetry is
\begin{equation}
S_{D_s^+ D_s^-} = \sin 2 \chi \,,
\end{equation}
which in the SM is of order $2 \lambda^2$.

\subsection{$\boldsymbol{b \rightarrow s \bar s s}$ with an extra up singlet}

\begin{figure}[htb]
\begin{center}
\epsfig{file=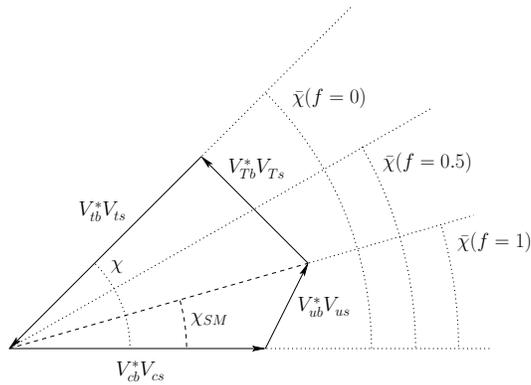,width=7cm,clip=}
\end{center}
\caption{Different values of $\bar \chi$ and its geometrical
meaning. The relative lengths of the sides of the quadrangle are illustrative.}
\label{fig:quad}
\end{figure}

In these models Eq.(\ref{ec:19}) is replaced by 
\begin{eqnarray}
A( \bar B_d^0 \rightarrow \phi \, \bar K^0) & = &
a(x_t) V_{tb} V_{ts}^* + a(x_T) V_{Tb} V_{Ts}^*\,,
\nonumber \\
A(B_d^0 \rightarrow \phi \, K^0) & = & a(x_t) V_{tb}^* V_{ts}
+ a(x_T) V_{Tb}^* V_{Ts} \,,
\label{ec:}
\end{eqnarray}
with $x_T = (\overline{m_T}/M_W)^2$,
due to the additional exchange of the $T$ quark. Similarly, Eq.~(\ref{ec:21}) is
generalised to
\begin{equation}
\lambda_{\phi K_S}= -e^{-i (2 \bar \beta + 2 \chi)}
\left( \frac{1 + f(x_T,x_t) V_{Tb} V_{Ts}^* / V_{tb} V_{ts}^*}{1 + f(x_T,x_t)
V_{Tb}^* V_{Ts} / V_{tb}^* V_{ts}} \right) \,,
\end{equation}
with $f(x_T,x_t) = a(x_T) /a(x_t)$.
Using the fact that  $2 |\lambda_{\phi K_S}|/(1+|\lambda_{\phi K_S}|^2) \simeq
1$ to a very good approximation, we obtain 
\begin{equation}
S_{\phi K_S} = \sin ( 2 \bar \beta +2 \bar \chi ) \,,
\label{ec:27}
\end{equation}
where the ``effective'' $\bar \chi$ for this process is defined as
\begin{equation}
\bar \chi = \chi -\frac{1}{2} \arg \left( \frac{1 + f(x_T,x_t) V_{Tb} V_{Ts}^*
/ V_{tb} V_{ts}^*}{1 + f(x_T,x_t) V_{Tb}^* V_{Ts} / V_{tb}^* V_{ts}} \right) \,.
\label{ec:28}
\end{equation}
The geometrical interpretation of the effective phase $\bar \chi$ can be seen
in Fig.~\ref{fig:quad}, for different values of $f$.
It is also useful to define $\chi_\mathrm{SM}$ as: 
\begin{equation}
\chi_\mathrm{SM} = \arg [ V_{cb} V_{cs}^* (V_{cs} V_{cb}^* + V_{us} V_{ub}^*) ]
 = \arg \left( 1+ \frac{V_{us} V_{ub}^*}{V_{cs} V_{cb}^*}\right)
\end{equation}
which equals $\chi$ in any model with $3\times 3$ unitarity. Since
$\sin \chi_\mathrm{SM} \leq |V_{us} V_{ub}|/|V_{cs} V_{cb}|$,
$\chi_\mathrm{SM} \sim \lambda^2$ even
when $3\times 3$ unitarity does not hold (see Fig.~\ref{fig:quad}).

\begin{figure}[htb]
\begin{center}
\begin{tabular}{ccc}
\epsfig{file=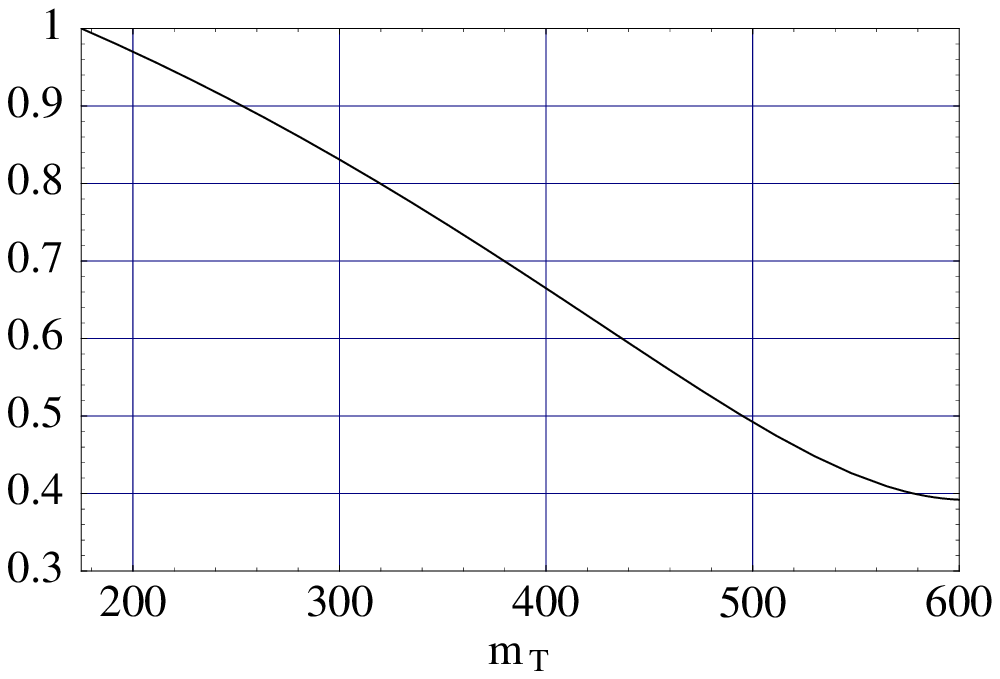,height=4cm,clip=} & ~~ &
\epsfig{file=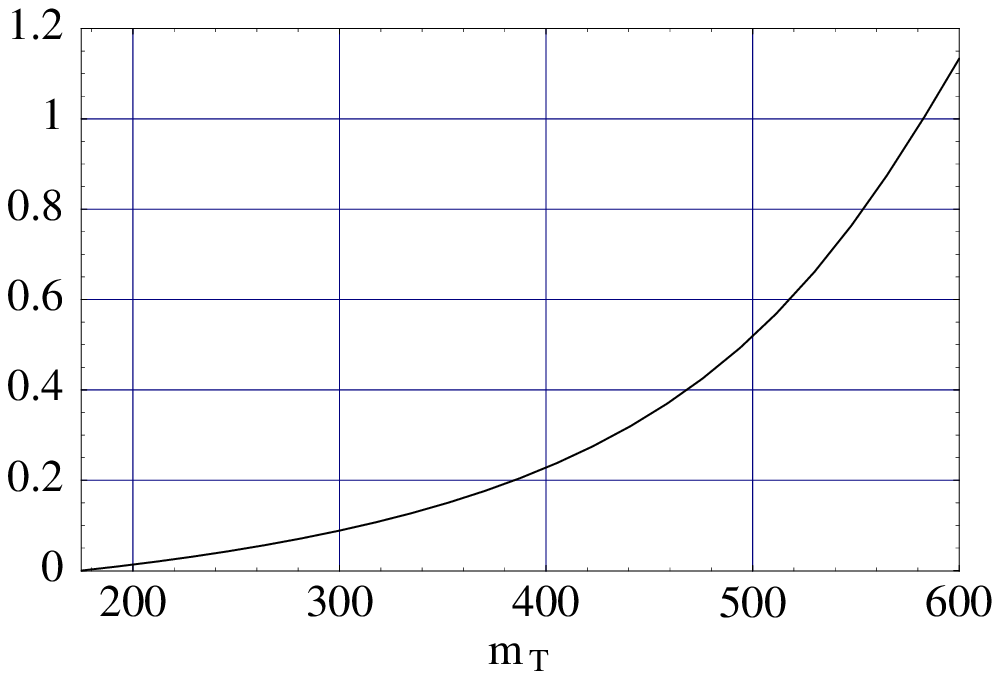,height=4cm,clip=} \\
(a) & & (b) 
\end{tabular}
\caption{Modulus (a) and argument (b) of $f$ as a function of $m_T$,
for fixed $x_t$.}
\label{fig:f}
\end{center}
\end{figure}

From Eq.~(\ref{ec:28})
it can be seen that in the limit $m_T = m_t$ the effective $\chi$
entering the CP asymmetry reduces to $\chi_\mathrm{SM}$,
\begin{eqnarray}
\lim_{m_T \rightarrow m_t} \bar \chi & = & \chi_\mathrm{SM} \,,
\end{eqnarray}
independently of the value of $\chi$. This ''screening'' property implies that,
despite the fact that the actual value of $\chi$ may be very different from the
SM prediction, the effective $\bar \chi$ that enters the CP
asymmetry is $O(\lambda^2)$ when $m_{T}$ tends to $m_{t}$. For larger $m_T$,
the degree of screening depends on the value of $f(x_T,x_t)$: for $f=0$ there
is no screening, and the screening is maximal for $f=1$. 
We calculate $a(x)$ using the QCD factorisation result of
Refs.~\cite{fact}, obtaining
\begin{eqnarray}
a(x) & = & -0.036880 - 0.012896 \, i- 0.005829 \, B_0(x) +0.004137 \, C_0(x)
 \nonumber \\
& & -0.000438 \, \tilde D_0(x) + 0.016376 \, E_0'(x) + 0.004074 \, \tilde E_0(x)
\,.
\end{eqnarray}
The Inami-Lim \cite{inami} functions $B_0$, $C_0$, etc. can be found in
Ref.~\cite{rev}. The
function $f(x_T,x_t)$ is plotted in Fig. \ref{fig:f} for fixed $x_t$.
The screening is important for 
low $m_T$, becoming milder as $m_T$ grows. In contrast, $\chi$ 
can be almost arbitrary for $m_T \sim m_t$, while its size is more
restricted for a heavier $T$, as can be observed in Fig.~\ref{fig:Xtt-X}.
With both effects working in opposite directions, we find that $S_{\phi K_S}$
is always inside the interval $[0.57,0.93]$, approaching the extremes for
heavier $T$. Since the screening is present in any $b \rightarrow s \bar s s$
transition, we expect a similar behaviour for all other
strong penguin dominated processes.

\subsection{$\boldsymbol{\bsmix}$ mixing with an extra up singlet}

\begin{figure}[htb]
\begin{center}
\epsfig{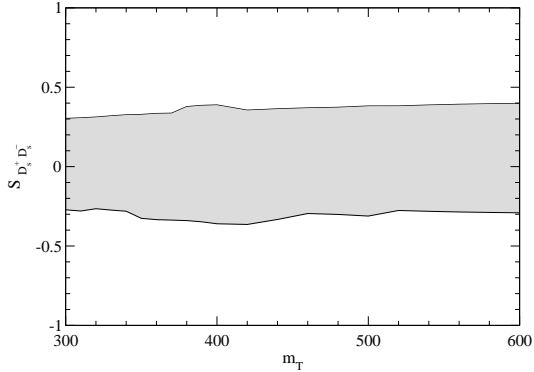} \\
\caption{Range of variation of the asymmetry $S_{D_s^+ D_s^-}$ (adapted from
Ref.~\cite{jaas}).}
\label{fig:aDsDs}
\end{center}
\end{figure}

With the addition of a $Q=2/3$ singlet,
the element $M_{12}$ of the $\bsmix$ mixing matrix can be
written as
\begin{equation}
M_{12}^{B_s} = K \sum_{i,j=t,T} (V_{is}^* V_{ib}) (V_{js}^* V_{jb}) 
S(x_i,x_j) = K S(x_t,x_t) |V_{ts}|^2 |V_{tb}|^2 r_s^2 e^{-2i\chi_\mathrm{eff}}
\,,
\label{ec:32}
\end{equation}
with $K$ a constant factor, $S$ the usual Inami-Lim box function
and
\begin{eqnarray}
r_s^2 e^{-2i\chi_\mathrm{eff}} & = & e^{-2i \chi} \left\{ \left[ 1+
\frac{S(x_t,x_T) V_{Ts}^* V_{Tb}}{S(x_t,x_t) V_{ts}^* V_{tb})} \right]^2 \right.
\nonumber \\
& & \left. + \left[ \frac{S(x_T,x_T)}{S(x_t,x_t)} -
\left( \frac{S(x_t,x_T)}{S(x_t,x_t)} \right)^2 \right] 
\left( \frac{V_{Ts}^* V_{Tb}}{V_{ts}^* V_{tb}} \right)^2 \right \} \,.
\label{BsMixing2}
\end{eqnarray}
The effective phase entering $\bsmix$ mixing is in this case
$\chi_\mathrm{eff}$, defined from the above equation.
In the
limit $x_{T}\rightarrow x_{t}$, the second term in the curly brackets goes to
zero and we get
\begin{equation}
\lim_{x_T \rightarrow x_t} \chi_\mathrm{eff} = \chi_\mathrm{SM}
\label{Screening2}
\end{equation}
as in the previous process. However,
in contrast with the function $f(x_T,x_t)$ which determines
the screening in the $b\rightarrow s \bar s s$ transitions, the ratio
$S(x_t,x_T)/S(x_t,x_t)$ in the first term of Eq.(\ref{BsMixing2}) is an
increasing function of $x_T$. This means that,
although for $x_T \rightarrow x_t$ the screening operates (as can be read from
Eq.~(\ref{Screening2})), for large $x_T$ we can have some enhancement of
$\chi_\mathrm{eff}$ with respect to $\chi$.
The range of variation of the asymmetry $S_{D_s^+ D_s^-} = \sin 2 
\chi_\mathrm{eff}$ is shown in Fig.~\ref{fig:aDsDs}. Although for heavier $T$
the allowed interval for $\chi$ is narrower, the enhancement above mentioned
makes the asymmetry be between $-0.4$ and $0.4$ for the $T$ masses
considered (this range of variation is quite different from the one predicted by
the SM). Such asymmetry could be easily be measured at LHCb, where
the expected precision in the $\psi \, \phi$ channel is around $0.066$ for one
year of running \cite{lhcb}.

\subsection{Unitarity and $\boldsymbol{\dmix}$ mixing}

The present experimental values of CKM matrix elements in the first row seem
to have a
discrepancy of $2.2-2.7$ standard deviations\cite{unit0}  with respect to the
SM unitarity prediction $|V_{ud}|^2 + |V_{us}|^2 + |V_{ub}|^2 =
1$.\footnote{Recent theoretical calculations \cite{unit1} and experimental
results \cite{unit2} would eliminate this discrepancy.} It is 
then worthwhile to question whether such apparent unitarity deviation
 could be
explained in scenarios with a large $\chi$, which also require a sizeable
breaking of $3 \times 3$ unitarity. We will show that this is not possible in
the minimal SM
extension studied here. In general, we have the inequality
\begin{equation}
|X_{uc}|^2 \leq (1-X_{uu}) (1-X_{cc}) \,,
\label{ec:desig2}
\end{equation}
but for only one extra singlet the equality holds. With $(1-X_{uu}) \sim
4 \times 10^{-3}$ (implying $|V_{u4}| \simeq 0.06$) from the apparent unitarity
deviation in the first row and
$(1-X_{cc}) \sim 10^{-3}$ in order to have large $\chi$, the FCN coupling
$X_{uc}$ would give a tree-level contribution to the $D^0$ mass difference
\cite{dlmD1,dlmD2} above the
present experimental limit $|\delta m_D| \leq 0.07$ ps$^{-1}$ \cite{pdb}.
In models with more than one extra singlet, the
equality in Eq.~(\ref{ec:desig2}) does not hold and this
argument is relaxed.

We also point out that, in this minimal extension with only one extra singlet,
$4 \times 4$ unitarity implies that in case $V_{Td}$ and $V_{u4}$ are both very
small $X_{ct}$ is also negligible. Since $V_{Td}$ must be small due
to constraints from $B$ oscillations (see for instance the matrices in
Eqs.~(\ref{ec:matr1}), (\ref{ec:matr2})), a large $\chi$ requires $V_{u4}$
not much smaller than $10^{-2}$. Therefore, it is expected that a large $\chi$
is associated with a $D^0$ mass difference not far from the present
experimental limit.

\section{Effects at high energy colliders}
\label{sec:5}

As implied by Eqs. (\ref{ec:chi-x}) and (\ref{ec:desig}), the fact of having a
phase $\chi \sim \lambda$ has consequences in some high energy processes: rare
top decays, $t \bar t$ production at $e^+ e^-$ collisions and the direct
production of a new quark at LHC.

\subsection{Top decays $\boldsymbol{t \to cZ}$}

Top FCN decays are extremely suppressed within the SM and hence they are a clear
signal of New Physics, if observed. In SM extensions with $Q=2/3$ singlets the
tree-level FCN couplings $X_{ut}$ and $X_{ct}$ can be large enough to yield
measurable top FCN interactions. These vertices lead to rare top decays
$t \to uZ, cZ$ and single top production in
the processes $gu, gc \to Zt$ (in hadron collisions) and
$e^+ e^- \to t \bar u, t \bar c$ (in $e^+ e^-$ annihilation), plus the charge
conjugate processes (see Ref.~\cite{review} for a review).
The best sensitivity to a $Ztc$ coupling is provided by top decays $t \to cZ$
at LHC. With a luminosity of
100 fb$^{-1}$, FCN couplings $|X_{ct}| \simeq
0.015$ can be observed with more than $5\,\sigma$ statistical significance
\cite{review}.
With a luminosity of 6000 fb$^{-1}$, achievable in one year with a
high luminosity upgrade \cite{lum},
a $3 \, \sigma$ significance can be obtained for $|X_{ct}| \simeq 0.0031$.
A moderately small phase, for instance $\chi \simeq 0.15$, requires
$\mathrm{Im} \; X_{ct} \simeq 0.006$, which would be observed with
more than $5 \, \sigma$ significance.

\subsection{$\boldsymbol{t \bar t}$ production in $\boldsymbol{e^+ e^-}$
collisions}

Top pair production at a 500 GeV linear collider will provide a precise
determination of
the $Ztt$ coupling through the measurement of the total $t \bar t$ cross section
and the forward-backward asymmetry. The accuracy of the measurement of $X_{tt}$
is mainly
limited by theoretical uncertainties in the prediction of the total cross
section. In order to determine the sensitivity to deviations of $X_{tt}$ from
unity, a Monte Carlo calculation of this process is necessary \cite{priv}.
The best results are obtained with beam polarisations $P_{e^+} = 0.6$,
$P_{e^-} = -0.8$. We assume that theoretical uncertainties in the total cross
section can be reduced to 1\% or below, and
a luminosity of 1000 fb$^{-1}$, which can be collected in three years of
running. For the SM value $X_{tt} = 1$ the
top pair production cross section is $\sigma = 47.9 \pm 0.5$ fb (including
theoretical and statistical uncertainties) and the
forward-backward asymmetry $A_\mathrm{FB} = -0.375 \pm 0.004$ (the error quoted
is only statistical).
For a phase $\chi \simeq 0.15$, $X_{tt}$ must be
typically around 0.96, yielding
$\sigma = 49.4 \pm 0.5$ fb, $A_\mathrm{FB} = -0.360 \pm 0.004$, which amount to
a combined $4.5 \, \sigma$ deviation with respect to the SM prediction. On the
other hand, if no deviations from the SM predictions are found, a bound $X_{tt}
\geq 0.985$ can be set with a 90\% CL, implying that $-0.12 \leq \chi \leq
0.14$, an indirect limit complementing the ones which will be previously
available from low energy processes.

\subsection{Direct production of $\boldsymbol{T \bar T}$ pairs in hadron
collisions}

The last (but obviously not least important) effect correlated with the
presence of a
phase $\chi \sim \lambda$ is the direct production of the new quark $T$. A 
sizeable deviation of $X_{tt}$ from unity is only possible if the new quark is
not very heavy, otherwise the contribution of the new quark to the
$T$ parameter, given by Eq. (\ref{ec:deltaT}),
would exceed present experimental limits.
 With the experimental value $\Delta T = -0.02 \pm 0.13$ and
admitting at most a $2 \, \sigma$ deviation, a coupling $X_{tt} \simeq 0.96$
(as required by $\chi \simeq 0.15$) is acceptable if the new quark has a mass
below
approximately 850 GeV. A new quark with this mass can be produced in pairs via
strong interactions, with a total tree-level cross section of 170 fb.
The observability of the new quark can be estimated as follows.
For $m_T= 850$ GeV, $X_{tt} = 0.96$ the new quark decays mainly to $W b$ and
$Z t$, with branching ratios $\mathrm{Br}(T \to W b) = 0.7$,
$\mathrm{Br}(T \to Z t) = 0.3$. This new quark could be easily seen in its
semileptonic decays $T \bar T \to l^\pm \nu jjjj$, being the total tree-level
cross section of the process
$q \bar q, g g \to T \bar T \to W^+ b W^- \bar b \to l^+ \nu jjjj$ (including
standard detector cuts) 5.5 fb (the same cross section for the final state
$l^- \nu jjjj$) \cite{priv}. The $Wjjjj$ background can be greatly
reduced with suitable cuts requiring that the events have a kinematics
compatible with $T \bar T$ production. The tree-level cross sections after cuts
for $l^+ \nu jjjj$ and $l^- \bar \nu jjjj$ are 75 fb and 45 fb, respectively,
calculated with {\tt VECBOS} \cite{vecbos}.
Taking into account only statistical uncertainties, with 100 fb$^{-1}$ the
$T \bar T$ signal could be observed with a significance of $10 \,
\sigma$.

\section{Concluding remarks}
\label{sec:6}

We have emphasised that a large value of $\chi$ requires physics beyond the SM,
in particular violations of $3 \times 3$ unitarity of the CKM matrix. It has
been shown that if this unitarity breaking arises from the presence of down-type
isosinglet quarks, $\chi$ is still constrained to be of order $\lambda^2$ due to
the constraint from the $b \to s l^+ l^-$ decay. On the contrary, it has been
pointed out that in the presence of up-type quark singlets a relatively large
value of $\chi$ can be obtained, without entering into conflict with present
experimental data.

The implications of a large $\chi$ have been analysed in the context of a
minimal
model with one $Q=2/3$ singlet. We have found that a large $\chi$ can lead
to moderate
departures of the SM approximate relation $S_{\phi K_S} \simeq S_{\psi K_S}$,
with $S_{\phi K_S}$ approximately in the interval $[0.57,0.93]$ (the precise
range also depends on hadronic matrix elements). On the other
hand, the effects on the CP asymmetry $S_{D_s^+ D_s^-}$ (and related channels)
are much larger, with these asymmetries ranging in the interval $[-0.4,0.4]$. 
These results must be compared with the ones for models with extra down
singlets, where large departures of $S_{\phi K_S} \simeq S_{\psi K_S}$ can be
accomodated \cite{phiKS} but $S_{D_s^+ D_s^-}$ is small and very
close to the SM range \cite{jaas}. Therefore, we can distinguish three possible
New Physics scenarios:
\begin{enumerate}
\item If a small departure in the relation $S_{\phi K_S} \simeq S_{\psi K_S}$
and a large (but within $[-0.4,0.4]$ approximately) $S_{D_s^+ D_s^-}$ are
found, they may suggest the presence of a new $Q=2/3$ singlet.

\item If a large departure in $S_{\phi K_S} \simeq S_{\psi K_S}$ is confirmed,
but with $S_{D_s^+ D_s^-}$ very small, it may indicate the presence of a
$Q=-1/3$ singlet.

\item In case that $S_{D_s^+ D_s^-}$ is found outside the interval
$[-0.4,0.4]$, or if a large departure in $S_{\phi K_S} \simeq S_{\psi K_S}$ and
a large $S_{D_s^+ D_s^-}$ are simultaneously found, they require the presence
of New Physics beyond these SM extensions with extra quark singlets, for
instance supersymmetric models \cite{susy}, which in principle could also
explain the discrepancies in the two previous scenarios.
\end{enumerate}

If New Physics hints are observed at B factories, its identification
may be possible at a large collider, perhaps with the direct production of the
new
particles. In the SM extensions with extra up-type singlets studied
we have found four correlated effects
which can be investigated at three different types of colliders: ({\em
i\/}) a large phase $\chi$ which has consequences on $B$ oscillation phenomena
at $B$ factories; ({\em ii\/}) a FCN coupling $X_{ct}$ which leads to top decays
$t \to c Z$ observable at LHC; ({\em iii\/}) a deviation of $X_{tt}$ from unity,
which can be measured in $t \bar t$ production at TESLA;
({\em iv\/}) The direct production of a new quark at LHC.
These associated effects, especially the discovery of the new particles, are
crucial to establish the origin of New Physics, if observed.

\vspace{1cm}
\noindent
{\Large \bf Acknowledgements}

\vspace{0.4cm} \noindent
This work has been supported by the European Community's Human Potential
Programme under contract HTRN--CT--2000--00149 Physics at Colliders and by FCT
through projects CERN/FIS/43793/2002 and CFIF--Plurianual (2/91) and by MECD
under FPA2002-00612. The work of
J.A.A.S. has been supported by FCT under grant SFRH/BPD/12603/2003. M.N.
acknowledges MECD for a fellowship. F.J.B. acknowledges the warm hospitality
during his stay at IST, Lisbon,where the major part of this work was done.

\end{document}